\documentclass[11pt]{article}

\newenvironment{descit}[1]{\begin{quote}\noindent\textit{#1}}{\end{quote}}

\ifx\undefined\psfig\else \fi

\edef\psfigRestoreAt{\catcode`@=\number\catcode`@\relax}
\catcode`\@=11\relax
\newwrite\@unused
\def\typeout#1{{\let\protect\string\immediate\write\@unused{#1}}}
\def\figurepath{./}

\def\@nnil{\@nil}
\def\@empty{}
\def\@psdonoop#1\@@#2#3{}
\def\@psdo#1:=#2\do#3{\edef\@psdotmp{#2}\ifx\@psdotmp\@empty \else
    \expandafter\@psdoloop#2,\@nil,\@nil\@@#1{#3}\fi}
\def\@psdoloop#1,#2,#3\@@#4#5{\def#4{#1}\ifx #4\@nnil \else
       #5\def#4{#2}\ifx #4\@nnil \else#5\@ipsdoloop #3\@@#4{#5}\fi\fi}
\def\@ipsdoloop#1,#2\@@#3#4{\def#3{#1}\ifx #3\@nnil 
       \let\@nextwhile=\@psdonoop \else
      #4\relax\let\@nextwhile=\@ipsdoloop\fi\@nextwhile#2\@@#3{#4}}
\def\@tpsdo#1:=#2\do#3{\xdef\@psdotmp{#2}\ifx\@psdotmp\@empty \else
    \@tpsdoloop#2\@nil\@nil\@@#1{#3}\fi}
\def\@tpsdoloop#1#2\@@#3#4{\def#3{#1}\ifx #3\@nnil 
       \let\@nextwhile=\@psdonoop \else
      #4\relax\let\@nextwhile=\@tpsdoloop\fi\@nextwhile#2\@@#3{#4}}
\newread\ps@stream
\newif\ifnot@eof       % continue looking for the bounding box?
\newif\if@noisy        % report what you're making?
\newif\if@atend        % %%BoundingBox: has (at end) specification
\newif\if@psfile       % does this look like a PostScript file?
{\catcode`\%=12\global\gdef\epsf@start{%!}}
\def\epsf@PS{PS}
\def\epsf@getbb#1{%
\openin\ps@stream=#1
\ifeof\ps@stream\typeout{Error, File #1 not found}\else
   {\not@eoftrue \chardef\other=12
    \def\do##1{\catcode`##1=\other}\dospecials \catcode`\ =10
    \loop
       \if@psfile
	  \read\ps@stream to \epsf@fileline
       \else{
	  \obeyspaces
          \read\ps@stream to \epsf@tmp\global\let\epsf@fileline\epsf@tmp}
       \fi
       \ifeof\ps@stream\not@eoffalse\else
       \if@psfile\else
       \expandafter\epsf@test\epsf@fileline:. \\%
       \fi
          \expandafter\epsf@aux\epsf@fileline:. \\%
       \fi
   \ifnot@eof\repeat
   }\closein\ps@stream\fi}%
\long\def\epsf@test#1#2#3:#4\\{\def\epsf@testit{#1#2}
			\ifx\epsf@testit\epsf@start\else
\typeout{Warning! File does not start with `\epsf@start'.  It may not be a PostScript file.}
			\fi
			\@psfiletrue} % don't test after 1st line
{\catcode`\%=12\global\let\epsf@percent=%\global\def\epsf@bblit{%BoundingBox}}
\long\def\epsf@aux#1#2:#3\\{\ifx#1\epsf@percent
   \def\epsf@testit{#2}\ifx\epsf@testit\epsf@bblit
	\@atendfalse
        \epsf@atend #3 . \\%
	\if@atend	
	   \if@verbose{
		\typeout{psfig: found `(atend)'; continuing search}
	   }\fi
        \else
        \epsf@grab #3 . . . \\%
        \not@eoffalse
        \global\no@bbfalse
        \fi
   \fi\fi}%
\def\epsf@grab #1 #2 #3 #4 #5\\{%
   \global\def\epsf@llx{#1}\ifx\epsf@llx\empty
      \epsf@grab #2 #3 #4 #5 .\\\else
   \global\def\epsf@lly{#2}%
   \global\def\epsf@urx{#3}\global\def\epsf@ury{#4}\fi}%
\def\epsf@atendlit{(atend)} 
\def\epsf@atend #1 #2 #3\\{%
   \def\epsf@tmp{#1}\ifx\epsf@tmp\empty
      \epsf@atend #2 #3 .\\\else
   \ifx\epsf@tmp\epsf@atendlit\@atendtrue\fi\fi}

\chardef\letter = 11
\chardef\other = 12

\newif \ifdebug %%% turn me on to see TeX hard at work ...
\newif\ifc@mpute %%% don't need to compute some values
\c@mputetrue % but assume that we do

\let\then = \relax
\def\r@dian{pt }
\let\r@dians = \r@dian
\let\dimensionless@nit = \r@dian
\let\dimensionless@nits = \dimensionless@nit
\def\internal@nit{sp }
\let\internal@nits = \internal@nit
\newif\ifstillc@nverging
\def \Mess@ge #1{\ifdebug \then \message {#1} \fi}

{ %%% Things that need abnormal catcodes %%%
	\catcode `\@ = \letter
	\gdef \nodimen {\expandafter \n@dimen \the \dimen}
	\gdef \term #1 #2 #3%
	       {\edef \t@ {\the #1}%%% freeze parameter 1 (count, by value)
		\edef \t@@ {\expandafter \n@dimen \the #2\r@dian}%
		\t@rm {\t@} {\t@@} {#3}%
	       }
	\gdef \t@rm #1 #2 #3%
	       {{%
		\count 0 = 0
		\dimen 0 = 1 \dimensionless@nit
		\dimen 2 = #2\relax
		\Mess@ge {Calculating term #1 of \nodimen 2}%
		\loop
		\ifnum	\count 0 < #1
		\then	\advance \count 0 by 1
			\Mess@ge {Iteration \the \count 0 \space}%
			\Multiply \dimen 0 by {\dimen 2}%
			\Mess@ge {After multiplication, term = \nodimen 0}%
			\Divide \dimen 0 by {\count 0}%
			\Mess@ge {After division, term = \nodimen 0}%
		\repeat
		\Mess@ge {Final value for term #1 of 
				\nodimen 2 \space is \nodimen 0}%
		\xdef \Term {#3 = \nodimen 0 \r@dians}%
		\aftergroup \Term
	       }}
	\catcode `\p = \other
	\catcode `\t = \other
	\gdef \n@dimen #1pt{#1} %%% throw away the ``pt''
}

\def \Divide #1by #2{\divide #1 by #2} %%% just a synonym

\def \Multiply #1by #2%%% allows division of a dimen by a dimen
       {{%%% should really freeze parameter 2 (dimen, passed by value)
	\count 0 = #1\relax
	\count 2 = #2\relax
	\count 4 = 65536
	\Mess@ge {Before scaling, count 0 = \the \count 0 \space and
			count 2 = \the \count 2}%
	\ifnum	\count 0 > 32767 %%% do our best to avoid overflow
	\then	\divide \count 0 by 4
		\divide \count 4 by 4
	\else	\ifnum	\count 0 < -32767
		\then	\divide \count 0 by 4
			\divide \count 4 by 4
		\else
		\fi
	\fi
	\ifnum	\count 2 > 32767 %%% while retaining reasonable accuracy
	\then	\divide \count 2 by 4
		\divide \count 4 by 4
	\else	\ifnum	\count 2 < -32767
		\then	\divide \count 2 by 4
			\divide \count 4 by 4
		\else
		\fi
	\fi
	\multiply \count 0 by \count 2
	\divide \count 0 by \count 4
	\xdef \product {#1 = \the \count 0 \internal@nits}%
	\aftergroup \product
       }}

\def\r@duce{\ifdim\dimen0 > 90\r@dian \then   % sin(x) = sin(180-x)
		\multiply\dimen0 by -1
		\advance\dimen0 by 180\r@dian
		\r@duce
	    \else \ifdim\dimen0 < -90\r@dian \then  % sin(x) = sin(360+x)
		\advance\dimen0 by 360\r@dian
		\r@duce
		\fi
	    \fi}

\def\Sine#1%
       {{%
	\dimen 0 = #1 \r@dian
	\r@duce
	\ifdim\dimen0 = -90\r@dian \then
	   \dimen4 = -1\r@dian
	   \c@mputefalse
	\fi
	\ifdim\dimen0 = 90\r@dian \then
	   \dimen4 = 1\r@dian
	   \c@mputefalse
	\fi
	\ifdim\dimen0 = 0\r@dian \then
	   \dimen4 = 0\r@dian
	   \c@mputefalse
	\fi
	\ifc@mpute \then
		\divide\dimen0 by 180
		\dimen0=3.141592654\dimen0
		\dimen 2 = 3.1415926535897963\r@dian %%% a well-known constant
		\divide\dimen 2 by 2 %%% we only deal with -pi/2 : pi/2
		\Mess@ge {Sin: calculating Sin of \nodimen 0}%
		\count 0 = 1 %%% see power-series expansion for sine
		\dimen 2 = 1 \r@dian %%% ditto
		\dimen 4 = 0 \r@dian %%% ditto
		\loop
			\ifnum	\dimen 2 = 0 %%% then we've done
			\then	\stillc@nvergingfalse 
			\else	\stillc@nvergingtrue
			\fi
			\ifstillc@nverging %%% then calculate next term
			\then	\term {\count 0} {\dimen 0} {\dimen 2}%
				\advance \count 0 by 2
				\count 2 = \count 0
				\divide \count 2 by 2
				\ifodd	\count 2 %%% signs alternate
				\then	\advance \dimen 4 by \dimen 2
				\else	\advance \dimen 4 by -\dimen 2
				\fi
		\repeat
	\fi		
			\xdef \sine {\nodimen 4}%
       }}

\def\Cosine#1{\ifx\sine\UnDefined\edef\Savesine{\relax}\else
		             \edef\Savesine{\sine}\fi
	{\dimen0=#1\r@dian\multiply\dimen0 by -1
	 \advance\dimen0 by 90\r@dian
	 \Sine{\nodimen 0}
	 \xdef\cosine{\sine}
	 \xdef\sine{\Savesine}}}	      
\def\psdraft{
	\def\@psdraft{0}
}
\def\psfull{
	\def\@psdraft{100}
}

\psfull

\newif\if@draftbox
\def\psnodraftbox{
	\@draftboxfalse
}
\@draftboxtrue

\newif\if@prologfile
\newif\if@postlogfile
\def\pssilent{
	\@noisyfalse
}
\def\psnoisy{
	\@noisytrue
}
\psnoisy
\newif\if@bbllx
\newif\if@bblly
\newif\if@bburx
\newif\if@bbury
\newif\if@height
\newif\if@width
\newif\if@rheight
\newif\if@rwidth
\newif\if@angle
\newif\if@clip
\newif\if@verbose
\newif\if@scale
\def\@p@@sclip#1{\@cliptrue}

\def\@p@@sfile#1{\def\@p@sfile{null}%
	        \openin1=#1
		\ifeof1\closein1%
		       \openin1=\figurepath#1
			\ifeof1\typeout{Error, File #1 not found}
			   \if@bbllx\if@bblly\if@bburx\if@bbury% added 6/91 Rob Russell
			      \def\@p@sfile{#1}%
			   \fi\fi\fi\fi
			\else\closein1
			    \edef\@p@sfile{\figurepath#1}%
                        \fi%
		 \else\closein1%
		       \def\@p@sfile{#1}%
		 \fi}
\def\@p@@sfigure#1{\def\@p@sfile{null}%
	        \openin1=#1
		\ifeof1\closein1%
		       \openin1=\figurepath#1
			\ifeof1\typeout{Error, File #1 not found}
			   \if@bbllx\if@bblly\if@bburx\if@bbury% added 6/91 Rob Russell
			      \def\@p@sfile{#1}%
			   \fi\fi\fi\fi
			\else\closein1
			    \def\@p@sfile{\figurepath#1}%
                        \fi%
		 \else\closein1%
		       \def\@p@sfile{#1}%
		 \fi}

\def\@p@@sbbllx#1{
		\@bbllxtrue
		\dimen100=#1
		\edef\@p@sbbllx{\number\dimen100}
}
\def\@p@@sbblly#1{
		\@bbllytrue
		\dimen100=#1
		\edef\@p@sbblly{\number\dimen100}
}
\def\@p@@sbburx#1{
		\@bburxtrue
		\dimen100=#1
		\edef\@p@sbburx{\number\dimen100}
}
\def\@p@@sbbury#1{
		\@bburytrue
		\dimen100=#1
		\edef\@p@sbbury{\number\dimen100}
}
\def\@p@@sheight#1{
		\@heighttrue
		\dimen100=#1
   		\edef\@p@sheight{\number\dimen100}
}
\def\@p@@swidth#1{
		\@widthtrue
		\dimen100=#1
		\edef\@p@swidth{\number\dimen100}
}
\def\@p@@srheight#1{
		\@rheighttrue
		\dimen100=#1
		\edef\@p@srheight{\number\dimen100}
}
\def\@p@@srwidth#1{
		\@rwidthtrue
		\dimen100=#1
		\edef\@p@srwidth{\number\dimen100}
}
\def\@p@@sangle#1{
		\@angletrue
		\edef\@p@sangle{#1} %\number\dimen100}
}
\def\@p@@ssilent#1{ 
		\@verbosefalse
}
\def\@p@@sscale#1{
		\def\@p@scale{#1}
		\@scaletrue
}
\def\@p@@sprolog#1{\@prologfiletrue\def\@prologfileval{#1}}
\def\@p@@spostlog#1{\@postlogfiletrue\def\@postlogfileval{#1}}
\def\@cs@name#1{\csname #1\endcsname}
\def\@setparms#1=#2,{\@cs@name{@p@@s#1}{#2}}
\def\ps@init@parms{
		\@bbllxfalse \@bbllyfalse
		\@bburxfalse \@bburyfalse
		\@heightfalse \@widthfalse
		\@rheightfalse \@rwidthfalse
		\@scalefalse
		\def\@p@sbbllx{}\def\@p@sbblly{}
		\def\@p@sbburx{}\def\@p@sbbury{}
		\def\@p@sheight{}\def\@p@swidth{}
		\def\@p@srheight{}\def\@p@srwidth{}
		\def\@p@sangle{0}
		\def\@p@sfile{}
		\def\@p@scost{10}
		\def\@sc{}
		\@prologfilefalse
		\@postlogfilefalse
		\@clipfalse
		\if@noisy
			\@verbosetrue
		\else
			\@verbosefalse
		\fi
}
\def\parse@ps@parms#1{
	 	\@psdo\@psfiga:=#1\do
		   {\expandafter\@setparms\@psfiga,}}
\newif\ifno@bb
\def\bb@missing{
	\if@verbose{
		\typeout{psfig: searching \@p@sfile \space  for bounding box}
	}\fi
	\no@bbtrue
	\epsf@getbb{\@p@sfile}
        \ifno@bb \else \bb@cull\epsf@llx\epsf@lly\epsf@urx\epsf@ury\fi
}	
\def\bb@cull#1#2#3#4{
	\dimen100=#1 bp\edef\@p@sbbllx{\number\dimen100}
	\dimen100=#2 bp\edef\@p@sbblly{\number\dimen100}
	\dimen100=#3 bp\edef\@p@sbburx{\number\dimen100}
	\dimen100=#4 bp\edef\@p@sbbury{\number\dimen100}
	\no@bbfalse
}

\newdimen\p@intvaluex
\newdimen\p@intvaluey
\newdimen\@ffsetvalue
\newdimen\x@ffsetvalue
\newdimen\y@ffsetvalue

\def\compute@offset#1#2{{\dimen0=#1 sp\dimen1=#2 sp
			\advance\dimen1 by -\dimen0
			\dimen1=\sine\dimen1
			\dimen0=\cosine\dimen1
			\ifdim\dimen0<0sp \dimen1=0sp \fi
			\global\@ffsetvalue=\dimen1}}

\def\rotate@#1#2{{\dimen0=#1 sp\dimen1=#2 sp
		  \global\p@intvaluex=\cosine\dimen0
		  \dimen3=\sine\dimen1
		  \global\advance\p@intvaluex by -\dimen3
		  \global\p@intvaluey=\sine\dimen0
		  \dimen3=\cosine\dimen1
		  \global\advance\p@intvaluey by \dimen3
		  }}
\def\compute@bb{
		\no@bbfalse
		\if@bbllx \else \no@bbtrue \fi
		\if@bblly \else \no@bbtrue \fi
		\if@bburx \else \no@bbtrue \fi
		\if@bbury \else \no@bbtrue \fi
		\ifno@bb \bb@missing \fi
		\ifno@bb \typeout{FATAL ERROR: no bb supplied or found}
			\no-bb-error
		\fi
		\if@angle 
			\Sine{\@p@sangle}\Cosine{\@p@sangle}
			\compute@offset{\@p@sbblly}{\@p@sbbury}
			\x@ffsetvalue=\@ffsetvalue
			\compute@offset{\@p@sbburx}{\@p@sbbllx}
			\y@ffsetvalue=\@ffsetvalue

			\rotate@{\@p@sbbllx}{\@p@sbblly}
			\advance\p@intvaluex by -\x@ffsetvalue
			\advance\p@intvaluey by -\y@ffsetvalue
			\edef\@p@sbbllx{\number\p@intvaluex}
			\edef\@p@sbblly{\number\p@intvaluey}

			\rotate@{\@p@sbburx}{\@p@sbbury}
			\advance\p@intvaluex by \x@ffsetvalue
			\advance\p@intvaluey by \y@ffsetvalue
			\edef\@p@sbburx{\number\p@intvaluex}
			\edef\@p@sbbury{\number\p@intvaluey}
			{
			 \count0=\@p@sbbllx \count1=\@p@sbblly
		 	 \count2=\@p@sbburx \count3=\@p@sbbury
			 \dimen0=\@p@sbbllx sp\dimen1=\@p@sbblly sp
		 	 \dimen2=\@p@sbburx sp\dimen3=\@p@sbbury sp
			 \dimen203=\dimen2 \advance\dimen203 by -\dimen0
			 \dimen204=\dimen3 \advance\dimen204 by -\dimen1
			 \ifdim\dimen203<0sp 
			      \count203=\count2 \count2=\count0 
			      \count0=\count203 
			      \global\edef\@p@sbbllx{\number\count0}
			      \global\edef\@p@sbburx{\number\count2}
			 \fi
			 \ifdim\dimen204<0sp 
			       \count204=\count3
			       \count3=\count1
			       \count1=\count204
			       \global\edef\@p@sbblly{\number\count1}
			       \global\edef\@p@sbbury{\number\count3}
			 \fi
			}
		\fi
		\count203=\@p@sbburx
		\count204=\@p@sbbury
		\advance\count203 by -\@p@sbbllx
		\advance\count204 by -\@p@sbblly
		\edef\@bbw{\number\count203}
		\edef\@bbh{\number\count204}
}
\def\in@hundreds#1#2#3{\count240=#2 \count241=#3
		     \count100=\count240	% 100 is first digit #2/#3
		     \divide\count100 by \count241
		     \count101=\count100
		     \multiply\count101 by \count241
		     \advance\count240 by -\count101
		     \multiply\count240 by 10
		     \count101=\count240	%101 is second digit of #2/#3
		     \divide\count101 by \count241
		     \count102=\count101
		     \multiply\count102 by \count241
		     \advance\count240 by -\count102
		     \multiply\count240 by 10
		     \count102=\count240	% 102 is the third digit
		     \divide\count102 by \count241
		     \count200=#1\count205=0
		     \count201=\count200
			\multiply\count201 by \count100
		 	\advance\count205 by \count201
		     \count201=\count200
			\divide\count201 by 10
			\multiply\count201 by \count101
			\advance\count205 by \count201
		     \count201=\count200
			\divide\count201 by 100
			\multiply\count201 by \count102
			\advance\count205 by \count201
		     \edef\@result{\number\count205}
}
\def\@ScaleInHundreds#1{
		\in@hundreds{#1}{\@p@scale}{100}
		\edef#1{\@result}
}
\def\compute@wfromh{
		\in@hundreds{\@p@sheight}{\@bbw}{\@bbh}
		\edef\@p@swidth{\@result}
}
\def\compute@hfromw{
		\in@hundreds{\@p@swidth}{\@bbh}{\@bbw}
		\edef\@p@sheight{\@result}
}
\def\compute@handw{
		\if@height 
			\if@width
			\else
				\compute@wfromh
			\fi
		\else 
			\if@width
				\compute@hfromw
			\else
				\edef\@p@sheight{\@bbh}
				\edef\@p@swidth{\@bbw}
			\fi
		\fi
}
\def\compute@resv{
		\if@rheight \else \edef\@p@srheight{\@p@sheight} \fi
		\if@rwidth \else \edef\@p@srwidth{\@p@swidth} \fi
}
\def\compute@sizes{
	\compute@bb
	\compute@handw
	\compute@resv
}
\def\psfig#1{\vbox {
	\ps@init@parms
	\parse@ps@parms{#1}
	\compute@sizes
	\if@scale
                \if@verbose
                        \typeout{psfig: scaling by \@p@scale}
                \fi
                \@ScaleInHundreds{\@p@swidth}
                \@ScaleInHundreds{\@p@sheight}
                \@ScaleInHundreds{\@p@srwidth}
                \@ScaleInHundreds{\@p@srheight}
        \fi
	\ifnum\@p@scost<\@psdraft{
		\if@verbose{
			\typeout{psfig: including \@p@sfile \space }
		}\fi
		\special{ps::[begin] 	\@p@swidth \space \@p@sheight \space
				\@p@sbbllx \space \@p@sbblly \space
				\@p@sbburx \space \@p@sbbury \space
				startTexFig \space }
		\if@angle
			\special {ps:: \@p@sangle \space rotate \space} 
		\fi
		\if@clip{
			\if@verbose{
				\typeout{(clip)}
			}\fi
			\special{ps:: doclip \space }
		}\fi
		\if@prologfile
		    \special{ps: plotfile \@prologfileval \space } \fi
		\special{ps: plotfile \@p@sfile \space }
		\if@postlogfile
		    \special{ps: plotfile \@postlogfileval \space } \fi
		\special{ps::[end] endTexFig \space }
		\vbox to \@p@srheight true sp{
			\hbox to \@p@srwidth true sp{
				\hss
			}
		\vss
		}
	}\else{
		\if@draftbox{		
			\hbox{\fbox{\vbox to \@p@srheight true sp{
			\vss
			\hbox to \@p@srwidth true sp{ \hss \@p@sfile \hss }
			\vss
			}}}
		}\else{
			\vbox to \@p@srheight true sp{
			\vss
			\hbox to \@p@srwidth true sp{\hss}
			\vss
			}
		}\fi

	}\fi
}}
\def\psglobal{\typeout{psfig: PSGLOBAL is OBSOLETE; use psprint -m instead}}
\psfigRestoreAt

\newif\ifpdf
\ifx\pdfoutput\undefined
  \pdffalse
\else
  \pdfoutput=1
  \pdftrue
\fi

\ifpdf
  \usepackage[pdftex]{graphicx}
  \usepackage[pdftex]{color}
  \DeclareGraphicsExtensions{.pdf,.png,.jpg}
\else
  \usepackage[dvips]{graphicx}
  \usepackage[dvips]{color}
  \DeclareGraphicsExtensions{.eps,.epsi,.ps}
\fi

\usepackage{times}

\def\midv{\mathop{\,|\,}}

\long\def\cbk#1{{\color{red}[CBK: #1]}}
\newlength\colwidth \setlength\colwidth{3.25in}

\title{Supporting Out-of-turn Interactions\\
in a Multimodal Web Interface}
\author{Atul Shenoy, Naren Ramakrishnan, \\
Manuel A. P\'{e}rez-Qui\~{n}ones, and Srinidhi Varadarajan\\
Department of Computer Science\\
Virginia Tech, VA 24061, USA\\
Contact email: naren@cs.vt.edu}
\date{}
\begin{document}

\maketitle
\begin{abstract}
\noindent
Multimodal interfaces are becoming increasingly important with the advent
of mobile devices, accessibility considerations, and novel software
technologies that combine diverse interaction media. This article
investigates systems support for web browsing in a multimodal interface.
Specifically, we outline the design and implementation of a software
framework that integrates hyperlink and speech modes of interaction. 
Instead of viewing speech as merely an alternative interaction medium, the
framework uses it to support out-of-turn interaction, providing a flexibility
of information access not possible with hyperlinks alone.
This approach enables the creation of websites that adapt to the needs of 
users, yet permits the designer fine-grained control over what interactions to 
support. Design methodology, implementation details, and two case studies 
are presented.
\end{abstract}

\noindent
{\bf Keywords:} Multimodal interfaces, web interaction on mobile devices,
dialog processing engines, mixed-initiative interaction.

%\tableofcontents
%\newpage
\section{Introduction}
\label{intro}
Computing power today is increasingly moving away from the desktop computer 
to mobile computing devices such as PDAs, tablet PCs, and 3G phones. 
While posing capacity limitations (e.g., screen real estate, memory), such 
devices also present possibilities for multimodal interaction via gestures, 
speech, and handwriting recognition. 

An area that is witnessing tremendous growth in multimodal interaction is web
browsing on mobile devices. Technologies such as SALT (Speech Application
Language Tags) and X+V (XHTML plus Voice)
are ushering in the speech-enabled 
web -- documents that can talk and listen rather than passively display content. The
maturing of commercial speech recognition engines~\cite{speech-mainstream} has been a key
factor in the emergence of this niche segment of multimodal browsing.

Speech as a mode of web interaction has become important for primarily two reasons. 
First, speech permits natural ways to perform certain types of tasks
and helps compensate for deficiencies 
in traditional hyperlink access (which can get cumbersome on small form-factor
devices). More importantly, speech-enabled
websites help improve accessibility for the more
than 40 million visually impaired people in the world today. As a result,
using speech leads to
the possibility of a conversational user interface~\cite{speechsurvey} that combines the expressive 
freedom of voice backed by the information bandwidth of a traditional browser. 

What exactly would we use a speech-enabled web interface for? The common use of
speech on a website is to support navigation of existing site structure via voice~\cite{vox}, in
other words as an alternative interaction medium. 
We posit that this is a rather limited viewpoint and that speech can actually be used 
to support new functionality at a website. In particular, we show how a multimodal
web interface can support a flexibility of information access not 
possible with hyperlinks alone.

\subsection*{Motivating Example}
Consider the following dialogs between an information seeker (Sallie) and
an automated political information system.

\begin{descit}{Dialog 1}
\vspace{-0.1in}
\begin{tabbing}
[x] \= abcdefab \= thiscanactuallybeamuchlongersentenceokay \kill
1 \> {\bf System:} \> Welcome. Are you looking for a Senator or a Representative?\\
2 \> {\bf Sallie:} \> Senator.\\
3 \> {\bf System:} \> Democrat or Republican or an Independent?\\
4 \> {\bf Sallie:} \> Republican.\\
5 \> {\bf System:} \> What State?\\
6 \> {\bf Sallie:} \> Minnesota.\\
7 \> {\bf System:} \> That would be Norm Coleman. First elected in 2002, Coleman ...\\
(conversation continues)
\end{tabbing}
\end{descit}

\begin{descit}{Dialog 2}
\vspace{-0.1in}
\begin{tabbing}
[x] \= abcdefab \= thiscanactuallybeamuchlongersentenceokay \kill
1 \> {\bf System:} \> Welcome. Are you looking for a Senator or a Representative?\\
2 \> {\bf Sallie:} \> Senator.\\
3 \> {\bf System:} \> Democrat or Republican or an Independent?\\
4 \> {\bf Sallie:} \> Not sure, but represents the state of Indiana.\\
5 \> {\bf System:} \> Well, then it is either a Democrat or a Republican, there
are no Independents from Indiana.\\
6 \> {\bf Sallie:} \> I see. Who is the Republican Senator?\\
7 \> {\bf System:} \> That would be Richard G. Lugar. First elected in 1976, Lugar ... \\
(conversation continues)
\end{tabbing}
\end{descit}

\noindent
It is helpful to contrast these dialogs from a conversational initiative standpoint.
In the first dialog, Sallie responds 
to the questions in the order they are posed by the system. Such a dialog is called a 
{\it fixed-initiative}
dialog as the initiative resides with the system at all times. 
The second dialog is system-initiated till Line 4,
where Sallie's input becomes unresponsive 
and provides some information that was 
not solicited. We say that Sallie has taken the initiative of conversation from the
system. Nevertheless, the conversation is not stalled, the system registers that
Sallie answered a different question than was asked, and refocuses the dialog
in Line 5 to the issue of party (this time, narrowing down the available options
from three to two). Sallie now responds to the initiative and the dialog progresses
to complete the specification of a political official. Such a conversation where
the two parties exchange initiative is called 
a {\it mixed-initiative interaction}~\cite{allen-aimag}.

\begin{figure}
\centering
\begin{center}
\begin{tabular}{cc}
\includegraphics[height=80mm]{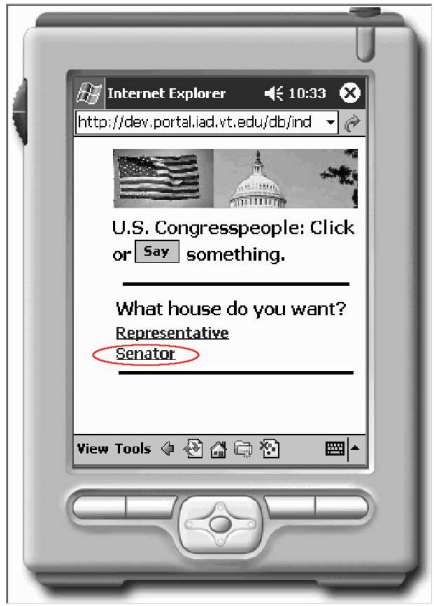}& \includegraphics[height=80mm]{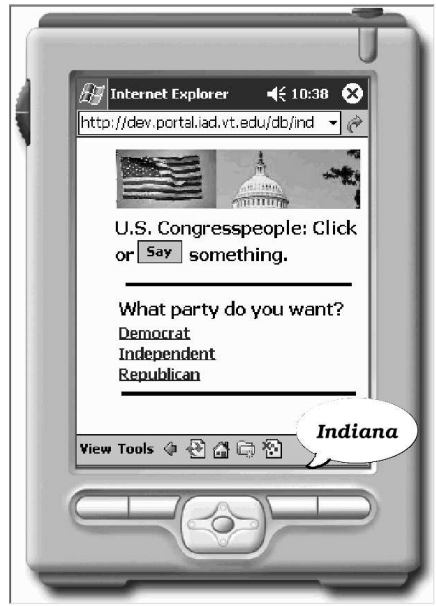}\\
(a) Sallie clicks on `Senator'. &(b) Sallie says `Indiana'. \\
\includegraphics[height=80mm]{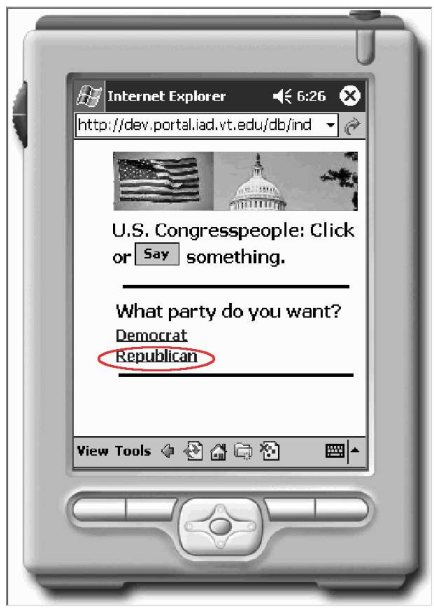}& \includegraphics[height=80mm]{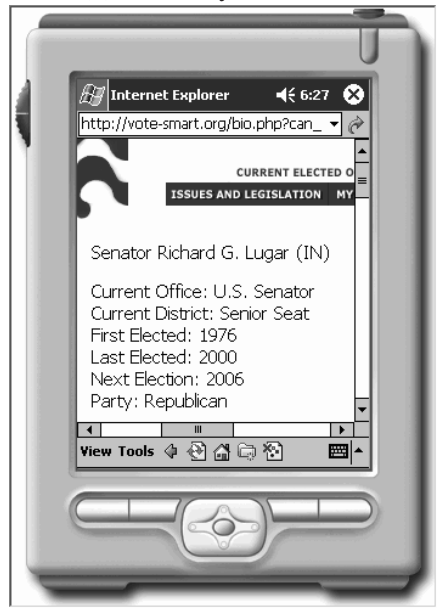}\\
(c) Sallie clicks on `Republican'. & (d) The dialog is complete. \\
\end{tabular}
\end{center}
\vspace{-0.2in}
\caption{A mixed-initiative interaction with a multimodal web interface.}
\vspace{-0.1in}
\label{teaser1}
\end{figure}

What would be required to support such a flexibility of interaction at a website?
It is clear that system-initiated modes of interaction are easiest to support and are
the most prevalent in web browsing today. For
instance, a webpage displaying a choice of hyperlinks presents such a view, so that 
clicking on a hyperlink corresponds to Sallie responding to the initiative. The reader
can verify that the first dialog above can be supported by a three-level tree-structured
HTML site presenting options for branch of congress, party, and state.
But how can we support the second dialog, allowing
Sallie to take the initiative at a website? This is where speech comes in. If 
Sallie can talk into the browser, she can provide unsolicited information using voice
when she is unable to make a choice among the presented hyperlinks. In addition,
if the system can process such an out-of-turn input, it can continue the progression 
of dialog and tailor future webpages so that they accurately reflect the information
gathered over the course of the interaction.
We have designed many such multimodal web interfaces, one example is shown in 
Fig.~\ref{teaser1}.

It is important to re-iterate that speech input is used here to provide a certain out-of-turn
interaction capability at a website. In other words, Sallie is not merely using voice
to answer the posed question (although she can do that too), but using it to 
specify unsolicited information. In the absence of such an out-of-turn facility,
the website designer would have to anticipate various user needs
and support all possible interaction sequences
directly in the HTML site structure (e.g., browse by branch-party-state order, browse
by branch-state-party order etc.), or provide a search facility as a method for pruning
web pages. The first solution is inelegant due to the mushrooming of choices, and the
second is not desirable either since search facilities usually terminate the dialog
and return a flat list of results. Out-of-turn interaction via speech does not
clutter the interface and provides a smooth continuation of the dialog.

We must point out that we are not supporting free-form
input of all kinds, only input pertaining to specification aspects that are not yet
solicited by the system. This can be viewed as akin to ``looking under the hood, and 
saying a hyperlink label that is deeper below.''

\section{System Design}
Supporting such a natural mode of interaction in a web interface is not an easy
undertaking. While technologies such as SALT and X+V enable the 
augmentation of speech into browsers, they either operate at a lower level of 
specification than the applications considered here, hence significantly increasing
programming effort, or are otherwise limited in their
expressive power for creating and managing dialogs~\cite{voxml-salt}. A multimodal web 
application must build on these technologies to provide flexible dialog capabilities.

Several considerations emerge in thinking about a software system design for
multimodal web interaction.
First, it is important to have uniform processing of hyperlink and voice interaction and,
when voice is used, to introduce minimal overhead in handling responsive versus
unsolicited input. Observe that hyperlink access can only be used to respond to the
initiative whereas voice input can be used both for responding
and for taking the initiative. Furthermore, a user may combine these modes
of initiative in a given utterance -- e.g., if the user speaks ``Republican Senator from
Minnesota'' at the outset, he is responding to the current solicitation as well
as providing two unsolicited pieces of information. Uniform processing of input modalities
irrespective of medium (hyperlink or voice) or initiative (responsive or unsolicited)
is thus important to support a seamless multimodal interface.
Second, it is beneficial to have a representation of the dialog
at all times, in order to determine how the user's input affects remaining
dialog options. For instance, in Line 5 of {\it Dialog 2} above, Independents are
removed as a possible party choice; in addition to pruning the hyperlink structure 
(shown in Fig.~\ref{teaser1}(c)), we must dynamically reconfigure the speech
recognizer to only await the remaining legal utterances. 
Third, it must be possible for the site designer to exert fine-grained control over 
what types of out-of-turn interactions are to be supported. 
And finally, it is desirable to be able to automatically re-engineer existing 
websites for multimodal out-of-turn interaction, without manual
configuration.

\begin{figure*}
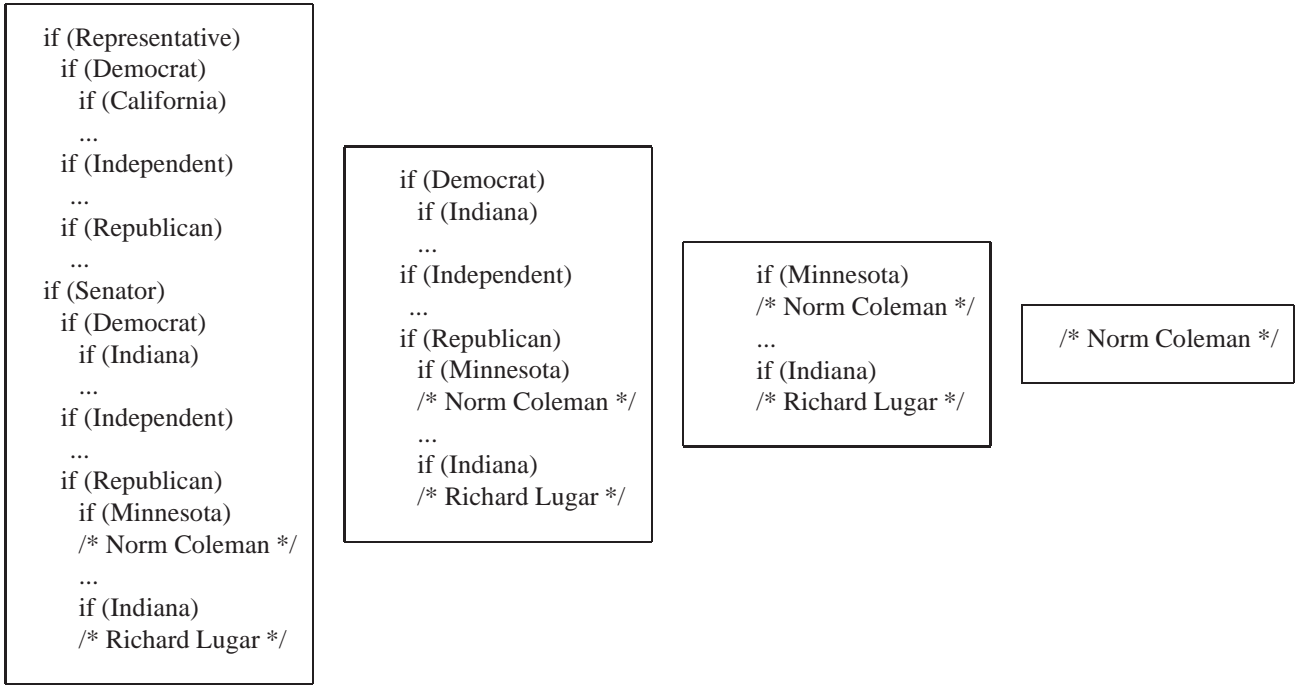

\centering
\begin{tabular}{cccc}
\begin{tabular}{|p{1.4in}|} \hline
\vspace{-0.2in} 
\small
\begin{tabbing}
\,\,\,\,\,if (Representative) \\ 
\,\,\,\,\,\,\,\,\,if (Democrat) \\ 
\,\,\,\,\,\,\,\,\,\,\,\,\,if (California) \\
\,\,\,\,\,\,\,\,\,\,\,\,\,... \\
\,\,\,\,\,\,\,\,\,if (Independent) \\ 
\,\,\,\,\,\,\,\,\,\,\,... \\
\,\,\,\,\,\,\,\,\,if (Republican) \\ 
\,\,\,\,\,\,\,\,\,\,\,... \\
\,\,\,\,\,if (Senator) \\ 
\,\,\,\,\,\,\,\,\,if (Democrat) \\ 
\,\,\,\,\,\,\,\,\,\,\,\,\,if (Indiana) \\
\,\,\,\,\,\,\,\,\,\,\,\,\,... \\
\,\,\,\,\,\,\,\,\,if (Independent) \\ 
\,\,\,\,\,\,\,\,\,\,\,... \\
\,\,\,\,\,\,\,\,\,if (Republican) \\ 
\,\,\,\,\,\,\,\,\,\,\,\,\,if (Minnesota) \\
\,\,\,\,\,\,\,\,\,\,\,\,\,/* Norm Coleman */ \\
\,\,\,\,\,\,\,\,\,\,\,\,\,... \\
\,\,\,\,\,\,\,\,\,\,\,\,\,if (Indiana) \\
\,\,\,\,\,\,\,\,\,\,\,\,\,/* Richard Lugar */ 
\end{tabbing}
\vspace{-0.2in}
\\\hline
\end{tabular}
&
\begin{tabular}{|p{1.4in}|} \hline
\vspace{-0.2in}
\small
\begin{tabbing}
\,\,\,\,\,\,\,\,\,if (Democrat) \\ 
\,\,\,\,\,\,\,\,\,\,\,\,\,if (Indiana) \\
\,\,\,\,\,\,\,\,\,\,\,\,\,... \\
\,\,\,\,\,\,\,\,\,if (Independent) \\ 
\,\,\,\,\,\,\,\,\,\,\,... \\
\,\,\,\,\,\,\,\,\,if (Republican) \\ 
\,\,\,\,\,\,\,\,\,\,\,\,\,if (Minnesota) \\
\,\,\,\,\,\,\,\,\,\,\,\,\,/* Norm Coleman */ \\
\,\,\,\,\,\,\,\,\,\,\,\,\,... \\
\,\,\,\,\,\,\,\,\,\,\,\,\,if (Indiana) \\
\,\,\,\,\,\,\,\,\,\,\,\,\,/* Richard Lugar */ 
\end{tabbing}
\vspace{-0.2in}
\\\hline
\end{tabular}
&
\begin{tabular}{|p{1.4in}|} \hline
\vspace{-0.2in}
\small
\begin{tabbing}
\,\,\,\,\,\,\,\,\,\,\,\,\,if (Minnesota) \\
\,\,\,\,\,\,\,\,\,\,\,\,\,/* Norm Coleman */ \\
\,\,\,\,\,\,\,\,\,\,\,\,\,... \\
\,\,\,\,\,\,\,\,\,\,\,\,\,if (Indiana) \\
\,\,\,\,\,\,\,\,\,\,\,\,\,/* Richard Lugar */ 
\end{tabbing}
\vspace{-0.2in}
\\\hline
\end{tabular}
&
\begin{tabular}{|p{0.05in}|} \hline
\vspace{-0.2in}
\small
\begin{tabbing}
\,\,\,\,\,/* Norm Coleman */ 
\end{tabbing}
\vspace{-0.2in}
\\\hline
\end{tabular}
\end{tabular}
\caption{Staging a system-initiated dialog using program transformations. 
The user specifies (`Senator,' `Republican,' `Minnesota'), in that order.}
\label{sens-staged1}
\vspace{-0.2in}
\end{figure*}

\begin{figure*}
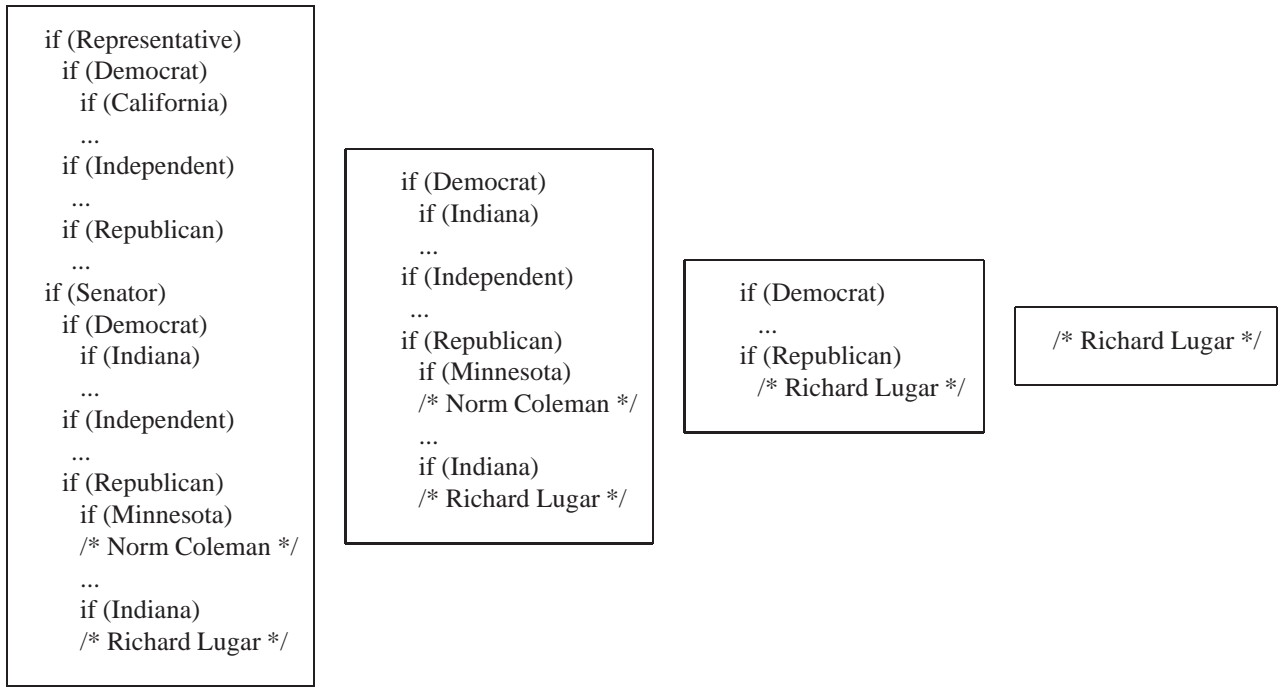

\centering
\begin{tabular}{cccc}
\begin{tabular}{|p{1.4in}|} \hline
\vspace{-0.2in} 
\small
\begin{tabbing}
\,\,\,\,\,if (Representative) \\ 
\,\,\,\,\,\,\,\,\,if (Democrat) \\ 
\,\,\,\,\,\,\,\,\,\,\,\,\,if (California) \\
\,\,\,\,\,\,\,\,\,\,\,\,\,... \\
\,\,\,\,\,\,\,\,\,if (Independent) \\ 
\,\,\,\,\,\,\,\,\,\,\,... \\
\,\,\,\,\,\,\,\,\,if (Republican) \\ 
\,\,\,\,\,\,\,\,\,\,\,... \\
\,\,\,\,\,if (Senator) \\ 
\,\,\,\,\,\,\,\,\,if (Democrat) \\ 
\,\,\,\,\,\,\,\,\,\,\,\,\,if (Indiana) \\
\,\,\,\,\,\,\,\,\,\,\,\,\,... \\
\,\,\,\,\,\,\,\,\,if (Independent) \\ 
\,\,\,\,\,\,\,\,\,\,\,... \\
\,\,\,\,\,\,\,\,\,if (Republican) \\ 
\,\,\,\,\,\,\,\,\,\,\,\,\,if (Minnesota) \\
\,\,\,\,\,\,\,\,\,\,\,\,\,/* Norm Coleman */ \\
\,\,\,\,\,\,\,\,\,\,\,\,\,... \\
\,\,\,\,\,\,\,\,\,\,\,\,\,if (Indiana) \\
\,\,\,\,\,\,\,\,\,\,\,\,\,/* Richard Lugar */ 
\end{tabbing}
\vspace{-0.2in}
\\\hline
\end{tabular}
&
\begin{tabular}{|p{1.4in}|} \hline
\vspace{-0.2in}
\small
\begin{tabbing}
\,\,\,\,\,\,\,\,\,if (Democrat) \\ 
\,\,\,\,\,\,\,\,\,\,\,\,\,if (Indiana) \\
\,\,\,\,\,\,\,\,\,\,\,\,\,... \\
\,\,\,\,\,\,\,\,\,if (Independent) \\ 
\,\,\,\,\,\,\,\,\,\,\,... \\
\,\,\,\,\,\,\,\,\,if (Republican) \\ 
\,\,\,\,\,\,\,\,\,\,\,\,\,if (Minnesota) \\
\,\,\,\,\,\,\,\,\,\,\,\,\,/* Norm Coleman */ \\
\,\,\,\,\,\,\,\,\,\,\,\,\,... \\
\,\,\,\,\,\,\,\,\,\,\,\,\,if (Indiana) \\
\,\,\,\,\,\,\,\,\,\,\,\,\,/* Richard Lugar */ 
\end{tabbing}
\vspace{-0.2in}
\\\hline
\end{tabular}
&
\begin{tabular}{|p{1.4in}|} \hline
\vspace{-0.2in}
\small
\begin{tabbing}
\,\,\,\,\,\,\,\,\,if (Democrat) \\ 
\,\,\,\,\,\,\,\,\,\,\,\,\,... \\
\,\,\,\,\,\,\,\,\,if (Republican) \\ 
\,\,\,\,\,\,\,\,\,\,\,\,\,/* Richard Lugar */ 
\end{tabbing}
\vspace{-0.2in}
\\\hline
\end{tabular}
&
\begin{tabular}{|p{0.05in}|} \hline
\vspace{-0.2in}
\small
\begin{tabbing}
\,\,\,\,\,/* Richard Lugar */ 
\end{tabbing}
\vspace{-0.2in}
\\\hline
\end{tabular}
\end{tabular}
\caption{Staging a mixed-initiative dialog using program transformations. 
The user specifies (`Senator,' `Indiana,' `Republican'), in that order.}
\label{sens-staged2}
\vspace{-0.2in}
\end{figure*}

\subsection{Dialog Representation}
We have designed a framework taking into account all these considerations~\cite{atul-thesis}.
It is based on {\it staging transformations}~\cite{staging} -- an approach that represents
dialogs by programs and uses program transformations to simplify them based on user
input. As an example, Fig.~\ref{sens-staged1} (left) depicts a representation of the
dialog from Section~\ref{intro} in a programmatic notation. You can see that the
tree-structured nature of the website is represented as a nested program of
conditionals, where each variable corresponds to a hyperlink that is present 
in the site. We can think of this program as being derived from a depth-first
traversal of the site. For {\it Dialog 1} of Section~\ref{intro},
the sequence of transformations in Fig.~\ref{sens-staged1} depicts what we 
want to happen. For {\it Dialog 2} of Section~\ref{intro}, 
the sequence of transformations in Fig.~\ref{sens-staged2} depicts what we 
want to happen. 

The first sequence of transformations corresponds to simply
interpreting the program in the order in which it is written. Thus, when
Sallie clicks on `Senator' she is specifying the values for the top-level of
nested conditionals (`Senator' is set to one, and `Representative' is set to zero).
This leads to a simplified program that now solicits for choice of party.
The sequence of Fig.~\ref{sens-staged2}, on the other hand,
corresponds to `jumping ahead' to nested segments and simplifying out
inner portions of the program before outer portions are even specified. This
transformation is well known to be {\it partial evaluation}, a technique
popular to compiler writers and implementors of programming systems~\cite{jones}.
In Fig.~\ref{sens-staged2} when the user
says `Indiana' at the second step, the program is partially evaluated with
respect to this variable (and variables for other states set to zero); the simplified
program continues to solicit for party, but one of the choices is pruned out
since it leads to a dead-end. Notice that
a given program when used with an interpreter corresponds
to a system-initiated dialog but morphs into a mixed-initiative dialog when
used with a partial evaluator! 

This is the essence of the staging transformation framework: using a program
to model the structure of the dialog and specifying a program transformer to
stage it. We write the first dialog as:
$${\frac{I}{\textrm{branch}\ \ \textrm{party}\ \ \textrm{state}}}$$
where the $I$ denotes an interpreter. Similarly, the second dialog is represented
as:
$${\frac{PE}{\textrm{branch}\ \ \textrm{party}\ \ \textrm{state}}}$$
where the $PE$ denotes a partial evaluator. An interpreter permits only
inputs that are responsive to the current solicitation and proceeds in a strict
sequential order; it results in the most restrictive dialog.
A PE, on the other hand, allows utterances of any combination of available
input slots in the dialog. It is the most flexible of stagers.

We will introduce a third stager, called a {\it curryer} ($C$) that permits utterance
of only valid prefixes of the input arguments. The dialog represented by
$${\frac{C}{\textrm{branch}\ \ \textrm{party}\ \ \textrm{state}}}$$
allows utterance of either `branch,' or (`branch,' `party'), or (`branch,'
`party,' `state'). In other words, if we are going to take the initiative at a
certain point, we must also answer the
currently posed question.

These stagers can be composed in a hierarchical fashion to yield dialogs 
comprised of smaller dialogs, or subdialogs. This allows us to make
fine-grained distinctions about the structure of dialogs and the range of
valid inputs. In this sense,
\[\frac{PE}{a\:b\:c\:d}\]
\vspace{-0.05in}
is not the same as
\vspace{-0.05in}
\[\frac{PE}{\frac{PE}{a\:b}\frac{PE}{c\:d}}\]
The former allows all $4!$ permutations of $\{a,b,c,d\}$ whereas
the latter precludes utterances such as $\prec c\:a\:b\:d\succ$.

As a practical example of our dialog
representation, consider a breakfast dialog involving specification of a 
\{eggs, coffee, bakery item\} tuple. The user can specify these
items in any order, but each item involves a second clarification aspect.
After the user has specified his choice of eggs,
a clarification of `how do you like
your eggs?' might be needed. Similarly, when the user is talking about
coffee, a clarification of `do you take cream and sugar?' might be required,
and so on. This form of mixing initiative is known as
{\it subdialog invocation}~\cite{allen-aimag}. The set of interaction sequences
that address this requirement can be represented as:
$${\frac{PE}{\frac{C}{e_1\:e_2}\frac{C}{c_1\:c_2}\frac{C}{b_1\:b_2}}}$$
where $e_1, e_2$ are egg specification aspects, $c_1, c_2$ support coffee
specification, and $b_1, b_2$ specify a bakery item.

The staging transformations framework also specifies a set of rules that
dictate how a (dialog, stager) pair is to be simplified based on user input.
Notice that this is not as straightforward as it looks as it might require
a global restructuring of the representation. Assume that we stage the breakfast
dialog using the interaction sequence $\prec c_1\:e_1\:c_2\:\cdots \succ$;
the occurrence of $e_1$ is invalid according to the dialog specification above,
but we will not know that such an input is arriving at the time we are processing
$c_1$. So in response to the input $c_1$, the dialog must be restructured as
follows: 
$${\frac{PE}{\frac{C}{e_1\:e_2}\frac{C}{c_1\:c_2}\frac{C}{b_1\:b_2}}}
\Longrightarrow
{\frac{C}{\frac{C}{c_2}\frac{PE}{\frac{C}{e_1\:e_2}\frac{C}{b_1\:b_2}}}}$$
By replacing the top-level PE stager with a C, it
becomes clear that the only legal input
now possible must have $c_2$. Once the coffee subdialog is completed, the top-level
stager will revert back to a PE. Such dialog restructurings are necessary 
if we are to remain faithful to the original specification. See~\cite{staging,atul-thesis}
for formal algorithms to perform such dialog restructurings.

\begin{figure*}
%\lstset{language=XML, showstringspaces=false, stringstyle=\ttfamily}
%\begin{lstlisting}[frame=trbl,breaklines=true,escapechar={!}]{final.xml}
\centering
\begin{tabular}{|p{5in}|} \hline
\vspace{-0.2in}
\begin{verbatim}
<?xml version="1.0" encoding="UTF-8"?>
<dialog-spec>
<dialog id="top" stager="pe" next="none" type="leaf">
        <dialog-item name="house" />
        <dialog-item name="party"/>
        <dialog-item name="state"/>
        <dialog-item name="seat"/>
        <dialog-item name="district"/>
</dialog>
</dialog-spec>
\end{verbatim}
\vspace{-0.2in}
\\\hline
\end{tabular}
\caption{DialogXML for the U.S. Congresspeople site. This dialog is staged by a partial
evaluator (PE) and consists of five specification aspects.}
\label{dialogxml1}
\end{figure*}

At this point, it must be clear that the staging transformation framework is
a powerful representational basis to design dialogs: it has a uniform vocabulary
for denoting specification aspects (e.g., each of the slots above could be filled
via hyperlink clicks or by voice) and the use of stagers helps us control the
mixing of initiative in a very precise manner. 

In order to make the staging notation
machine-readable, we have defined an XML representation of dialogs called
DialogXML. Fig.~\ref{dialogxml1} depicts a minimal DialogXML specification
for the politicians example. DialogXML provides
elements for defining the slots associated with a dialog, the textual prompts
associated with each slot (not shown in Fig.~\ref{dialogxml1}), the accompanying vocal prompts
and any tapering of them over the course of
interaction (also not shown), confirmatory characteristics (whether the user's response needs
confirmation), and constructs for combining basic dialog elements
to create complex dialogs.
More details about DialogXML and the possible legal specifications are available 
in~\cite{atul-thesis}. While DialogXML borrows ideas from some tags in the VoiceXML
standard~\cite{voicexml}, the structure of the DialogXML document
is more closely modeled after the idea of stagers\footnote{It has been recently
brought to our attention that there is a similar technology with the same 
name~\cite{otherdialogxml}. This work, however, is an extension of the VoiceXML standard
for voice interfaces and does not address multimodal interaction.}.

\begin{figure}[t]
  \begin{center}
    \includegraphics[height=70mm, width=170mm]{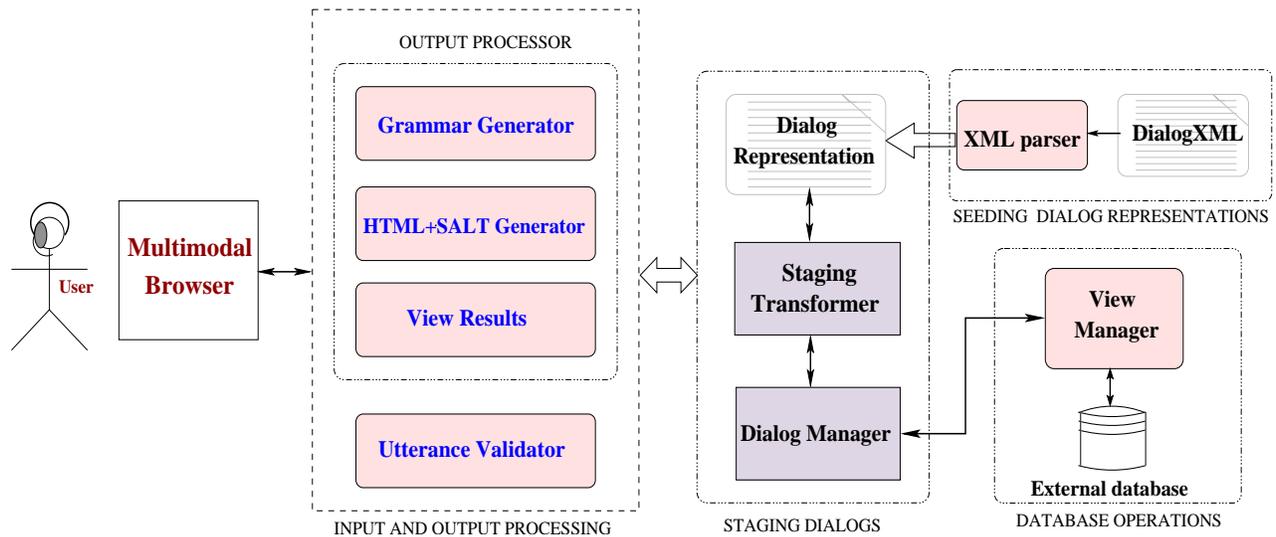}
    \caption{Multimodal out-of-turn interaction framework architecture.}
    \label{arch}
\vspace{-0.3in}
  \end{center}
\end{figure}

\subsection{Site Creation and Content Generation}
Having such a representation of the dialog is only the first step. We must create a site
to reflect the underlying structure of the dialog and initiate speech processing to
recognize the legal utterances as specified in the stager markup. Based on user input,
we must simplify the dialog and present personalized content, including facilities
for continuing the dialog.

An integrated
software framework for this purpose is shown in Fig.~\ref{arch}. Operationally it
can be divided into four modules: (i) seeding dialog representations, (ii)
staging dialogs, (iii) input and output processing, and (iv) database connectivity.
The idea of seeding dialog representations is to take a dialog specified in our
DialogXML notation and create an internal representation, suitable for staging.
The staging transformation framework is then used to handle dialog management.
The embodiment of these dialogs must contend with voice realizations, hence
a significant portion of the framework is devoted to generating grammars for
speech output and validating voice input. 
The framework currently uses the
SALT and Speech Recognition Grammars standards to handle voice interaction.
Finally, the database operations module manages and streamlines
the delivery of web content. Every website is organized as a database that
the user initially selects for exploration via out-of-turn interaction.
Each record in the database identifies a unique interaction sequence
leading to a leaf web page (e.g., Senator Norm Coleman is identified by
a record that describes his political affiliations and other addressable
attributes).

When the interaction begins, the 
dialog manager uses the parsed DialogXML and metadata from the database to 
initialize the representation.  The dialog manager must then decide 
what content to display on the page,
the items to offer the user for selection, and the speech prompts to
play for the current slot. 
In addition, the dialog manager must determine what aspects the user may specify
out-of-turn. Through an analysis of the dialog representation and a set of SQL queries,
it determines this information and generates a HTML page that contains
relevant SALT XML objects and references to a suitably generated SRGS grammar.
The grammar identifies all the legally speakable utterances for the particular
page.

User input from both voice and hyperlinks is uniformly handled by the Utterance
Validator module of the system. Upon receiving an utterance from the user, the module
first determines whether the utterance contains fillings for multiple slots, and whether
the utterance is valid. If part or whole of the utterance was invalid, then the
system accepts the valid utterances and rejects the invalid utterances.
An appropriate prompt is displayed and played to the user. Having tokenized the user's
utterance into its constituent fillings, the dialog manager then calls the
staging transformer with the values for the fillings in the order
they were received.  After the representation is simplified, the dialog
manager applies a suite of {\it dialog motivators}~\cite{construct-algebra} (discussed
below) to the 
dialog. If the dialog is not completed, content creation and speech grammar 
generation resumes.

The system is implemented as a Java web application using JSPs and Servlets. The 
application runs inside the Tomcat servlet engine. The system uses the Apache 
HTTPd web-server with a WARP connector to connect to the Tomcat servlet 
engine, and functions like a proxy-server. A PostgreSQL database server serves 
the example databases we use in this article. The web-application connects to the 
database server using the Java Database Connectivity (JDBC) API. It uses a meta-data API 
to learn about the structure of the data present in the database. An SQL query initially
helps compute a VIEW that serves as the starting point for the dialog. This VIEW is
used for all future interactions and helps reinforce that a user is always working
with a personalized `view' of the information space. The use of VIEWs
can be used to increase system efficiency as they can be shared across many users. 
We tested the system using the Microsoft SALT plug-in for the Internet 
Explorer 6.0 browser on a system running Windows XP.

\subsection{Dialog Motivators}

The only aspect of the architecture to be covered are the dialog motivators and the
grammar generators. Dialog motivators
are useful nuggets of processing that help streamline the dialog at every user
utterance. We use four main motivators:
\begin{enumerate}
\item {\bf complete-dialog:} This motivator decides if a dialog is complete. A dialog is 
complete if a unique record in the database VIEW being used has been identified, or if 
there are no more items left to solicit input for in the dialog. In such cases, the
unnecessary slots are removed from the representation.

\item {\bf prune-dialog:} This motivator decides if the internal dialog representation can be 
pruned as a result of the previous utterance by the user. For example, in the case 
of a pizza dialog, if the user specifies a size of `small,' and
only pepperoni pizzas available in small size, there is 
no longer a need to ask the user for a topping. Thus the topping slot can be 
automatically filled with `pepperoni' and removed from the dialog representation. 
While the current implementation does not provide the user with notification when a dialog 
is pruned in this fashion, such user-feedback is being considered for future work.

Notice that {\bf complete-dialog} is a specialization of {\bf prune-dialog}.

\item {\bf confirm-dialog:} This motivator applies if the item has been designated in the
DialogXML as one for which confirmation must be sought. In a real application, confirmation 
would be sought for utterances that have been recognized with a low value of confidence; 
however SALT does not provide us with the hooks to learn about confidence values 
of recognized utterances, hence we specify the need for confirmation in the 
DialogXML markup.
\item {\bf collect-results:} This motivator applies if the user explicitly requested 
(via a `Show me results' utterance)
that the dialog be terminated in order to view a flat listing (of the relevant remaining
records).
\end{enumerate}

\subsection{Grammar Generation}
Grammar generation proceeds in a straightforward manner except for a careful
division of labor between the browser, utterance validator, and the embedded
speech grammars themselves. For instance, 
the system generates JavaScript to handle some types of interaction on the client 
side within the browser itself. Confirmation of the user's utterance, 
`What may I say?,' and `Show me something else' type of questions are examples of
interactions handled by JavaScript. The embedded SRGS grammars are used for
encapsulating site-specific logic
and are faithful to the $C$ and $I$ stager specifications.
Generating grammar fragments corresponding to the $PE$ stager will result in
an exponential enumeration of utterance possibilities, so we use a less
restrictive grammar and allow the invalid utterances to be caught by the
Utterance Validator instead.

\section{Example Applications}
Two applications have been created using the out-of-turn interaction framework presented
above. The first is an interface to the Project VoteSmart website (http://www.vote-smart.org)
and an example interaction has been already described in Fig.~\ref{teaser1}. This
interface provides details on about 540 politicians comprising the U.S. Congress.

\begin{figure}
\centering
\begin{tabular}{cc}
\includegraphics[height=100mm]{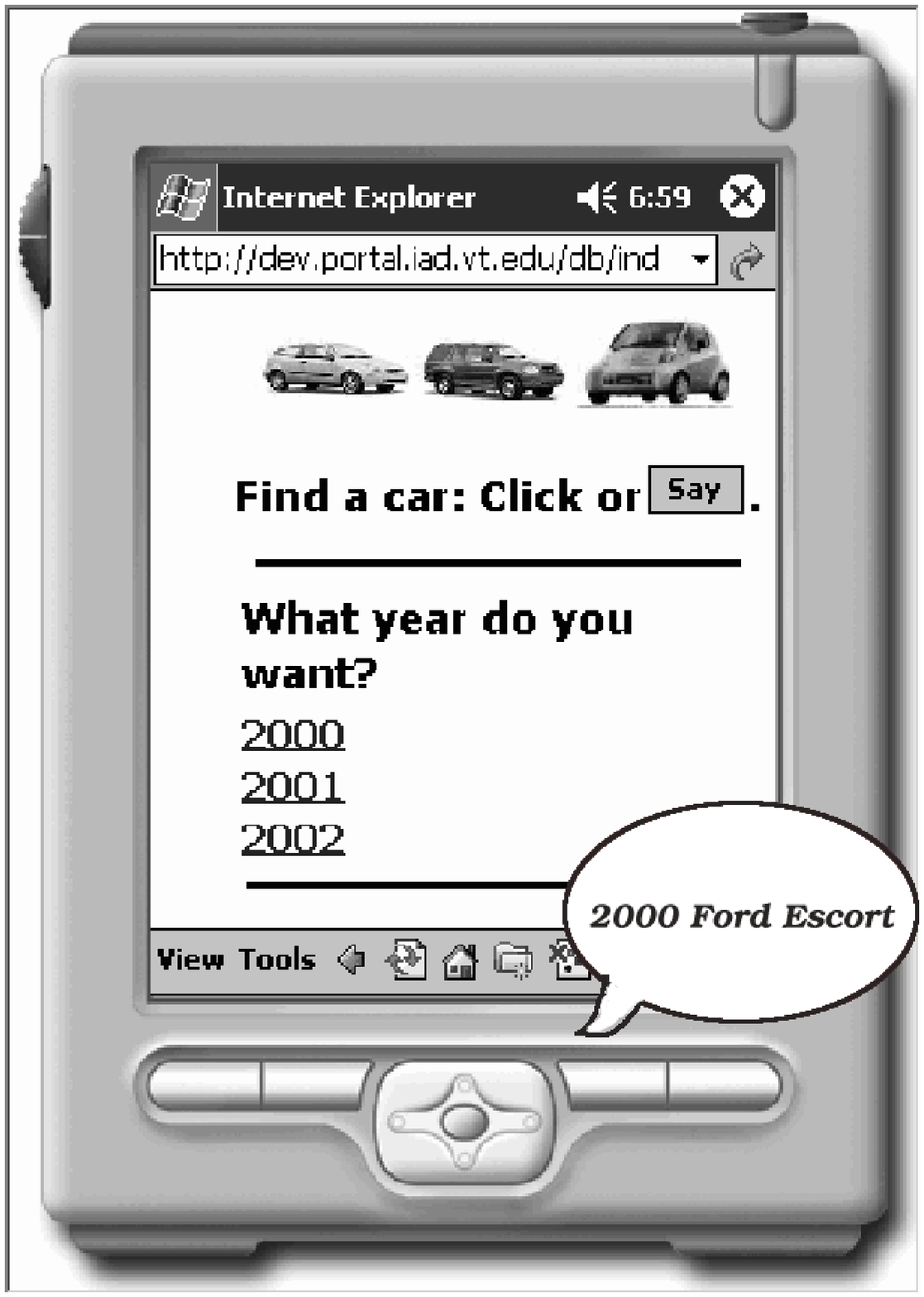} & \includegraphics[height=100mm]{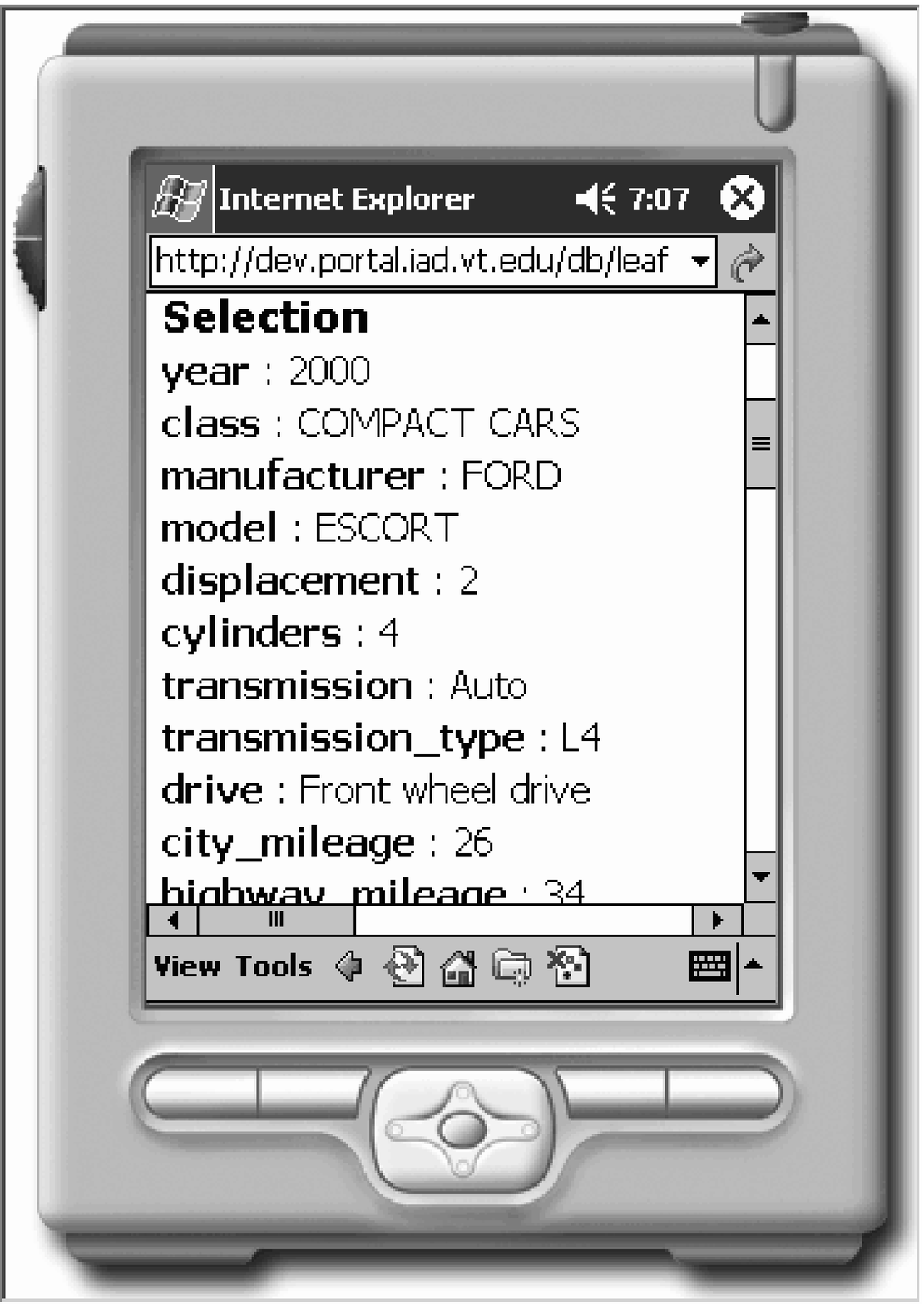} \\
(a) An out-of-turn utterance. &(b) The dialog is complete. \\
\end{tabular}
\caption{An interaction sequence using the multimodal web interface to the Fuel Economy Guide.}
\label{teaser2}
\end{figure}

The second application is an interface to the fuel economy guide at the environmental
protection agency (EPA -- http://www.fueleconomy.gov). Th EPA provides raw data on
fuel economy statistics about cars available in the United States in a comma separated
format (CSV). For this article, we downloaded and reformatted data from the past three
years (2000, 2001, and 2002) and loaded it into a PostgreSQL database. The dataset
has a total of 2641 records, which translates to approximately 880 records. Upto
26 different attributes can be specified in any interaction with this database.
We organized a dialog around three subdialogs. The first is an engine subdialog
which solicits (fuel type, information about whether the engine is a gas
guzzler, and if it is equipped with a turbo charger and/or a super charger). Another
subdialog solicits information about the transmission (whether it is automatic
or manual) and the drive (4 wheel or all wheel). The main specification
aspects for the car (year, manufacturer, model, number) are included in
another subdialog. While there are other ways to organize this information, we initialized
the dialog representation as:
\[\frac{PE}{\frac{PE}{\textrm{year}\:\:\textrm{class}\:\:\textrm{maker}\:\:\textrm{model}}\frac{C}{\textrm{fuel}\:\:\textrm{gas}\:\:\textrm{super charged?}\:\:\textrm{turbo charged?}}\frac{C}{\textrm{transmission}\:\:\textrm{drive}}}\]
An example interaction is shown in Fig.~\ref{teaser2}. 
It depicts an expert user who knows exactly what he wants, and as a result 
does not need to engage in a dialog with the system. In a single utterance, he specifies 
three pieces of information that uniquely identify a car in the database. The 
dialog manager has applied the {\bf prune-dialog} motivator to the items in 
the remaining dialog as these specification aspects are no longer necessary. The system 
redirects the user to a leaf page, where the user is able to see information about 
Ford Escort cars manufactured in 2000. This example also shows how the user 
is able to specify multiple utterances while interacting with the system.

We hasten to add that, for ease of presentation, the screenshots in this article were taken on a Microsoft PocketPC
simulator. As of this writing, PocketPC does not support SALT and these case studies were
actually tested using Microsoft Internet Explorer with the SALT plugin.

\section{Discussion}
The above applications have highlighted the key features of our software framework.
The staging transformers have been primarily responsible for dialog control, specifying
what the user may say at any given time. The dialog manager has streamlined the dialog, 
pruning it when necessary, and triggering the appropriate actions. Our modularized
implementation approach makes it easy to construct speech-enabled interfaces 
to database-driven sites. For want of space,
we have not demonstrated several other features of our system such
as user response confirmation, tapered prompting, and results collection.

This work helps demonstrate the viability of our view of the speech-enabled web --
namely, that of a flexible dialog between the user and the system which allows the user
to take the initiative in controlling the flow of the dialog. Rosenfeld et 
al.~\cite{rosenfeld-olsen} have argued that speech interfaces will become increasingly 
ubiquitous and will be able to support smaller form-factors without comprising 
usability. The applications presented here validate this viewpoint and help
illustrate the importance of using voice to supplement interaction in mobile devices.

It is helpful to contrast our representation-based approach with other ways of
specifying dialogs, notably VoiceXML. While they share some similarities, 
our DialogXML notation is purely
declarative and captures only the structure of the dialog. Control is implicitly
specified using program transformations, which makes the process of dialog specification
less cumbersome for the designer. Furthermore, while VoiceXML permits
mixed-initiative dialog sequences too, it does so more as a result of how its form
interpretation algorithm (FIA) is organized. Using a combination of program
transformation constructs and hierarchically composed dialogs, we are able to
specify the nature of out-of-turn interaction in a manner not precisely
expressable in VoiceXML (see~\cite{miimc} for more details). 

The successful implementation of a dialog-based system~\cite{elephant,jupiter} requires 
many more facets
such as language understanding,
task modeling, intention recognition, and plan management,
which are beyond the scope of this work.
We are now exploring several directions such as natural language speech input and
extending the specification capability of DialogXML. We are also conducting usability
studies for our multimodal interfaces and carefully assessing the role of speech as an
out-of-turn interaction medium. Especially important is addressing the veritable
`how do users know what to say?' problem~\cite{hdukwts} for 
multimodal web browsing.

This work is an initial exploration into the use of multimodal interfaces to websites.
As the use of browsers that support technologies such as SALT, X+V grows,
the importance of software frameworks to support multimodal interaction will only
increase.

\section*{Acknowledgements}
This work is supported in part by US National Science Foundation grants IIS-0049075, 
IIS-0136182, and by a grant from IBM to explore the use of VoiceXML within their 
WebSphere product. We acknowledge the helpful contributions of Michael Narayan (who
designed the formal rules underlying staging transformations), Robert Capra (for input
on speech-enabled interfaces), and Saverio Perugini (who implemented a toolbar version
of out-of-turn interfaces~\cite{itpro}). 
%Thanks also to Mill Mountain for supplying coffee.

\bibliographystyle{plain}
\bibliography{paper}

\end{document}